\documentclass[11pt,letterpaper]{JHEP3}

\usepackage{epsfig,multicol,bbm}
\usepackage{amsfonts}
\usepackage{amsmath}
\usepackage{mqcommands}

\newcommand{\reef}[1]{(\ref{#1})}
\renewcommand{\eqref}[1]{(\ref{#1})}
\newcommand{\php}{\phi_\omega}
\newcommand{\phm}{\phi_{-\omega}}

\title{Transport coefficients, membrane couplings and universality at extremality}

\author{Miguel F. Paulos$^{a}$\\
$^a$ {\it Department of Applied Mathematics and Theoretical
    Physics, Cambridge CB3 0WA, UK}\\

\vskip .5cm

{\rm E-mail:}\ \ {\
m.f.paulos@damtp.cam.ac.uk}}

\abstract{We present an efficient method for computing the zero frequency limit of transport coefficients in strongly coupled field theories described holographically by higher derivative gravity theories. Hydrodynamic parameters such as shear viscosity and conductivity can be obtained by computing residues of poles of the off-shell lagrangian density. We clarify in which sense these coefficients can be thought of as effective couplings at the horizon, and present analytic, Wald-like formulae for the shear viscosity and conductivity in a large class of general higher derivative lagrangians. We show how to apply our methods to systems at zero temperature but finite chemical potential. Our results imply that such theories satisfy $\eta/s=1/4\pi$ universally in the Einstein-Maxwell sector. Likewise, the zero frequency limit of the real part of the conductivity for such systems is shown to be universally zero, and we conjecture that higher derivative corrections in this sector do not modify this result to all orders in perturbation theory.}


\preprint{DAMTP-2009-70}

\begin{document}{\vskip 1cm}

\section{Introduction}

In recent years, with the AdS/CFT correspondence \cite{Maldacena}\cite{Magoo} extensively tested, there has been increasing interest in applying it as a tool to study strongly coupled systems. The case where the dual field theory has conformal symmetry is of particular interest as it could provide information on real-world systems such as the quark-gluon plasma and materials at quantum critical points. For recent reviews and further references on these topics see for instance \cite{HartnollLectures}\cite{GubserReview1}\cite{GubserReview2}.

According to the AdS/CFT dictionary, a conformal field theory at finite temperature is dual to a black hole living in an asymptotically AdS space. The thermodynamics of the dual field theory then correspond to the thermodynamics of the black hole. It can also be shown that small hydrodynamic fluctuations around thermodynamical equilibrium map onto small perturbations of the gravitational background in a precise fashion \cite{Minwalla}.

A hydrodynamic description is relevant at wavelengths and frequencies that are large compared to some typical microscopic scale of the theory. In this approach one describes the system by a set of transport coefficients. Determining these quantities from first principles is in general quite difficult. Typically we are interested in transport of conserved charges in the field theory, and in this case Kubo formulas can be used to read off the coefficients from low frequency poles in the thermal retarded Green's functions of conserved currents. This requires a study of real-time physics at finite temperature. While this can be achieved with some difficulty at weak coupling, at strong coupling the situation is thornier, as lattice methods are intrinsically euclidean. For some recent results on the shear viscosity of Yang-Mills theory plasmas see for instance \cite{Meyer}.

In contrast, in a holographic context these computations turn out to be remarkably simple. The prescription for calculating thermal retarded Green's functions was first set out by \cite{Policastro1}\cite{Policastro2}\cite{Policastro3}, and later more formally justified in \cite{Starinets},\cite{Skenderis}. Using this approach, all hydrodynamic transport coefficients up to second order have been computed in a large class of theories, as well as their first string theoretical corrections \cite{BuchelViscosity}\cite{Benincasa}\cite{RelaxationTime}\cite{QuantumEtaS}\cite{BoostInv}\cite{Universal}\cite{AnindaQM}\cite{Beyond}.

In a beautiful paper, the authors of \cite{LiuMembrane} showed that there is a very close link between the transport properties of the black hole ``stretched horizon'' in the membrane paradigm \cite{Membrane}, and those of the dual field theory. In this picture, the Green's function is essentially given by the canonical momentum associated to the bulk perturbation that sources the boundary conserved current. While the value of the canonical momentum at the horizon is fixed by regularity, generically there is a radial flow to the boundary. In the hydrodynamic limit this flow can be trivial, and in this case the properties of the field theory plasma coincide with those of the membrane. This is exactly what happens for the shear viscosity, from which the celebrated result $\eta/s=1/4\pi$ \cite{KSS} can be understood as emerging from the universal properties of black hole horizons. Universal properties of correlators in the hydrodynamic limit had been noticed already in \cite{BuchelUniversal}.

Extension of these methods for the computation of the shear viscosity in higher derivative theories have been given in \cite{Chemical},\cite{Banerjee}. The approaches of those papers are somewhat complementary. Whereas the first focuses on the generalization of the canonical momentum method, the second proposes to rewrite the effective action as a two derivative theory. The problem with the latter is that the overall coefficient of the effective action has to be fixed by hand, and it is not clear what its value should be in a general higher derivative theory (e.g. with derivatives of curvatures). 
Even though the two approaches should be in principle equivalent, both procedures rely on massaging the effective action into a desired form, which for general higher derivative theories can become quite cumbersome. Furthermore the treatment of possible contributions from boundary terms is not done in a systematic fashion.

In both these methods it is clear that generically transport coefficients associated to massless modes are given by effective couplings at the horizon. In this paper we show that there is a quick, efficient procedure for extracting these couplings from the lagrangian density. By evaluating the lagrangian on an off-shell perturbation, it will generically develop a pole at the horizon. The residue of the pole is precisely the desired transport coefficient up to a known factor. Boundary terms cannot contribute as in general they yield higher order singularities. The single pole behaviour can be traced back to horizon regularity of the generalized canonical momentum in the hydrodynamical approximation.
This procedure for computing transport coefficients, which we name the pole method, works for any higher derivative theory and reproduces all the previously known results in the literature in a simple fashion.

The idea of effective horizon couplings giving transport properties was pioneered in \cite{BrusteinViscosity},\cite{BrusteinEntropy}, where a specific covariant formula for the horizon graviton coupling was proposed, and therefore for the shear viscosity. However, it has been found in \cite{Banerjee} that this formula does not yield the correct result for the well known $C^4$ correction \cite{BuchelViscosity}. In this paper we clarify this issue, by carefully finding an analytic expression for membrane couplings, finding disagreement with the proposal of \cite{BrusteinViscosity}. Specifically we find formulae for both the shear viscosity and DC conductivity in an uncharged background for a wide class of lagrangians. These expressions match all previous calculations in the literature. 

Recently there has been interest in studying the transport properties for backgrounds at zero temperature but finite chemical potential \cite{LiuAdS2},\cite{Leigh},\cite{AnindaExtreme}. The relevant backgrounds describe a flow from the boundary field theory to an IR CFT, itself described by an $AdS_2$ factor in the near-horizon geometry . We show that the shear viscosity to entropy density ratio may be computed solely within the near horizon geometry, and therefore in Einstein gravity we deduce the universality of $\eta/s=1/4\pi$ at extremality. We show how to apply the pole method in this case, as well as giving a simplified analytic formula for the shear viscosity.

The computation of the conductivity is more interesting, as it exhibits a flow brought about by the background charge. As shown in \cite{Leigh}, in the IR CFT the real part of the conductivity scales like $\omega^2$ in the hydrodynamic limit. We give evidence that this result is actually valid generically for any higher derivative theory. The RG flow from the IR to the boundary theory cannot affect this result, which leads us to conclude that the low frequency limit of the conductivity of the dual theory is zero to all orders in the higher derivative expansion in the gravity-gauge sector. This is equivalently described as the non-renormalization of the conformal dimension of the specific scalar in IR CFT associated to charge transport.

The structure of this paper is as follows. In section 2 we describe the basic setup of our calculations for field theories at finite temperature.
The following section clarifies the notion of effective horizon couplings and shows that the generalized canonical momentum method is generally applicable.
Section 4 introduces our pole method for computing transport coefficients and some practical examples. In section 5 we derive analytic formulae for the shear viscosity and conductivity for a vast class of theories and compare them with previous proposals in the literature. Section 6 is devoted to the study of systems at zero temperature but finite chemical potential. We show that our methods are still applicable there and as an immediate corollary prove universality of $\eta/s=1/4\pi$ in two derivative theories. We also study the low frequency limit of the real part of the conductivity in general higher derivative theories and find evidence that it is zero to all orders in perturbation theory. We finish this paper with a discussion and future prospects. Some technical points are presented in the appendices.

\section{General Setup}

We are interested in computing transport coefficients for field theories at finite temperature living in $d$ flat dimensions. We assume that, in what concerns thermal properties and hydrodynamics, these theories admit an effective holographic description at strong coupling by a dual $(d+1)$-dimensional gravitational background of the form:
\bea
ds^2&=& g_{ab}dx^{a}dx^{b}=\frac {L^2}{z}\, e^{2g(z)} dz^2+g_{\mu\nu}dx^{\mu}dx^{\nu}\nonumber \\
g_{\mu\nu}&=&-z\,e^{2f(z)} dt^2+e^{2\rho(z)} dx^i dx_i \label{bg}.
\eea
The functions $f,g,\rho$ are regular everywhere in the bulk, which means that there is a horizon located at $z=0$. Henceforth we will denote quantities evaluated at the horizon by an index `$0$'.
The field theory lives at the boundary of the space at $z=1$ and is at temperature\footnote{The temperature can be obtained by fixing the periodicity of the Wick-rotated time coordinate as to eliminate the conical singularity at $z=0$.}
\begin{eqnarray}
T=\frac{1}{4\pi L} \exp(f_0-g_0).
\end{eqnarray}
It will be convenient to define the volume of the horizon as given by $V\equiv e^{(d-1)\rho_0}$.

We shall consider perturbations of the above background described by a massless scalar field $\phi(z,x^\mu)$. According to gauge-gravity duality, such a field sources an operator $\mathcal O$ in the dual field theory via a coupling $\int d^dx\mathcal O(x^\mu) \phi_B(x^\mu)$, where $\phi_{B}$ denotes the boundary value of the perturbation. 

From the field theory perspective, we have simply some source coupling to the operator $\mathcal O$. For small enough fields, and going to momentum space, linearized theory then determines the expectation value of $\mathcal O$ as
\be
\langle \mathcal O(k) \rangle=G_R(k)\phi_{B}(k).
\ee
In the low frequency limit and at zero spatial momentum one expects $\langle \mathcal O(\omega)\rangle=i\xi \omega \phi$. Therefore, to obtain $\xi$ we can use the Kubo formula
\be
\xi=\lim_{\omega \to 0} \frac 1{\omega}\, \mbox{Im}\, G_R(\omega,\mbf k=0). \label{xiformula}
\ee
Finding the transport coefficient $\xi$ then amounts to computing the retarded Green's function in the small frequency limit and at zero spatial momentum. At weak coupling it is possible to directly compute this Green's function in the field theory. However, at strong coupling it is simpler to perform this computation holographically.

\section{Retarded Green's functions in the hydrodynamic approximation}

In this section we show how to holographically compute thermal retarded Green's functions. We start by briefly reviewing the canonical momentum method of \cite{LiuMembrane} and then show how to generalize it for higher derivative theories, including a careful treatment of possible boundary terms. We study the regularity conditions at the horizon, and show precisely how the transport coefficient $\xi$ is related to an effective coupling at the horizon, along the lines of \cite{BrusteinViscosity}.

\subsection{Two derivative theory}

The procedure that must be taken to compute retarded Green's functions via gauge-gravity duality is by now relatively well understood.
Consider adding to the background \reef{bg} a small perturbation described by an action:
\bea
S_{\phi}^{(2)}=-\frac 12 \int d^{d} x\, dz\, \frac{ \sqrt{-g}}{\kappa}\,(\nabla \phi)^2.
\eea
We now go to momentum space, and set the spatial momentum to zero. This is accomplished by choosing the ansatz
\be
\phi(z,x)=\int \frac{d^dk}{(2\pi)^d} \phi_\omega(z)\delta(\mbf k) e^{i k_\mu x^\mu}
\ee
with $k_\mu=(-\omega,\mbf k)$. The action becomes
\bea
S_{\phi}^{(2)}&=& \int \prod_{i=1}^{d-1} dx^i \int \frac{d\omega}{2\pi} \int_{0}^1 dz \,\frac{ -\sqrt{-g}}{2\kappa} \left(g^{zz}\phi'_\omega(z)\phi'_{-\omega}(z)+g^{tt}\omega^2 \phi_{\omega}(z)\phi_{-\omega}(z)\right)\label{lowaction},
\eea
Following \cite{LiuMembrane}, the retarded Green's function is given by the formula
\bea
G_R(k)= \lim_{z\to 1} \frac{\pi_\omega(z)}{\phi_\omega(z)} \label{Kubo}
\eea
where $\pi_\omega(z)$ is the radial canonical momentum associated with $\phi_\omega(z)$:
\bea
\pi_\omega(z)=\frac{\delta S^{(2)}_\phi}{\delta (\partial_z \phi)}=-\frac{\sqrt{-g}}{\kappa} g^{zz} \partial_z \phi_\omega(z).
\eea
Now consider the small frequency limit where we take $\omega/T\to 0$ keeping $\pi_\omega(z)$ fixed. In these conditions the radial evolution of both canonical momentum and $\omega \phi$ is trivial \cite{LiuMembrane} (as we will show in the next section), and so we are free to evaluate the ratio in \reef{Kubo} at any radius we wish. We will see that it is convenient to choose the horizon for this so that,
\be
\xi= \lim_{\omega\to 0} \frac{\pi_\omega(0)}{i\omega\, \phi_0},
\ee
where we have used $\omega \phi(z)=\omega\, \phi_0$.

\label{gencanmom}
\subsection{The generalized canonical momentum}
We would like to generalize this procedure to theories containing higher derivatives. Such theories can arise for instance by considering string theory corrections to the supergravity action. It is implicit that there should be some small parameters controlling the size of such corrections. Then the higher derivative terms are perturbatively small, and the equations of motion can always be recast as two derivative equations. In this paper we shall always assume that whenever there higher derivative terms present, they are always parameterically smaller than the two derivative terms.

We will continue to restrict ourselves to massless fields, and work in Fourier space at zero spatial momenta. In this case, the most general quadratic action can always be written as:
\bea
S^{(2)}_{\phi}=\int \prod_{i=1}^{d-1} dx^i \int \frac{d\omega}{2\pi}\left(S_{(z)}+S_{(t)}+S_B\right) \label{ac1}.
\eea
The first piece in the above is the ``radial'' action. This contains all the terms in the action without any time derivatives, or equivalently the action that one obtains by setting $\omega=0$. The second piece contains the remaining terms, and by time reversal invariance it is necessarily proportional to $\omega^2$. Finally $S_B$ contains boundary terms necessary to make the variational problem well defined\footnote{For higher derivative theories, this has to be done perturbatively in the parameters controlling the higher derivative terms. See e.g. \cite{Benincasa} for an example on how this is done in practice.}. We will consider radial actions of the form
\bea
S_{(z)}&=& \int_{0}^1 dz \left(\sum_{n,m\geq 0}A_{n,m}(z)\phi_{\omega}^{(n+1)}(z)\phi_{-\omega}^{(m+1)}(z)\right), \label{ac2}
\eea
where $\phi_{\omega}^{(p)}(z)$ indicates $(\partial_z)^{p} \phi_\omega(z)$.
We should emphasize that in picking the specific form of the action \reef{ac1},\reef{ac2}, the only constraint is that the perturbation $\phi$ should be massless. Under these conditions, it is always possible to integrate by parts so that we get an action of this form. We will see that in applications these manipulations will be unneccessary.

For an action given by \reef{ac1},\reef{ac2} the boundary action necessarily contains only three types of terms. The first is simply $B_0(z) \phi^2$. This type of boundary terms does not contribute to the imaginary part of the Green's function, so we will not be concerned with them. Then there are boundary terms proportional to $\omega^2$, which in the small frequency limit do not contribute.
Finally there are terms of the form $B_{n,m}(z)\phi_\omega^{(n+1)}\phi_{-\omega}^{(m+1)}$. These terms must always have at least one derivative on each $\phi_\omega(z)$. To see this, consider for instance the following term in the radial action,
\be
A_{1,2}(z)\phi_{\omega}'(z)\phi_{-\omega}''(z)
\ee
Variation leads to boundary terms %
\be
\partial_z \left(A_{1,2}(z)\delta\phi_{\omega}(z)\phi_{-\omega}''(z)+A_{1,2}(z)\phi'_{\omega}(z)\delta\phi_{-\omega}'(z)\right)
\ee
We may set the first term to zero, but we must have a boundary term of the type $\phi' \phi'$ to cancel the second term. We see that no undifferentiated $\phi$ appears. It is clear that considering terms in the radial action with more derivatives will lead to similar results.
To sum up our description of the boundary action, we conclude that, in the low frequency limit, the only possible relevant boundary terms are such that they only contain differentiated fields.

Under these conditions we define the generalized canonical momentum:
\be
\Pi_\omega(z)\equiv \frac{\delta S_{z}}{\delta (\partial_z \phi_{-\omega})} \label{canmom}
\ee
To apply this definition, one can simply treat $\partial_z \phi_{-\omega}$ as a new field $\psi(z)$ and then treat $\phi,\psi$ as independent fields. The canonical momentum is then the functional derivative with respect to the $\psi(z)$ field. There is no ambiguity in this definition, as $\phi_{\omega}(z)$ always appears differentiated in the radial action. With this definition in hand, we can integrate by parts so that the radial action becomes
\be
S_{(z)}=\int_{0}^1 dz \,
\left(\frac 12 \Pi_\omega(z)\phi_{-\omega}'(z)\right) \label{SwithPi}
\ee
This rewriting will produce some extra boundary terms which are covered by our previous discussion.
Then it is clear that the equation of motion derived from the total action \reef{ac1} is of the form:
\be
\partial_z \Pi_\omega(z)=\omega^2 F(z,\phi,\phi',...).
\ee
The Green's function is once again given by the value of the on-shell action:
\be
G_R(\omega)=\lim_{z\to 1} \frac{ \Pi_\omega(z)}{\phi_{\omega}(z)}+\mbox{Boundary terms}. \label{GRHD}
\ee
In the small frequency limit, the equation of motion implies $\partial_z \Pi_\omega\simeq 0$. On the other hand, notice that at strictly $\omega=0$, a constant field $\phi(z)=\phi_0$ must be a solution, and therefore $\partial_z \phi(z)=\mathcal O(\omega)$. In this way we have shown that just as in the two derivative case we have a trivial flow from horizon to boundary, and so we are free to evaluate the ratio in \reef{GRHD} at the horizon as before. 

One needs to consider as well the possible contributions of boundary terms. As discussed above, a typical term is of the form $\partial_z [B(z) \phi^{n+1}(z) \phi^{m+1}(z)]$. This form can be traced back to the way in which we picked our radial action. But any differentiated $\phi(z)$ is necessarily $\mathcal O(\omega)$ as we've seen. Then automatically these boundary terms are all of order $\omega^2$, and therefore cannot contribute to the imaginary part of the Green's function in the low frequency limit.

Using \reef{xiformula} and \reef{GRHD} we conclude that
\be
\xi= \lim_{\omega\to 0} \frac{\Pi_\omega(0)}{i\omega \phi_0}. \label{xifrompi}
\ee
This shows that for a large class of higher derivative theories, the imaginary part of the retarded Green's functions at zero spatial momentum and small frequency is still given by the canonical momentum term. This generalizes the results of \cite{Chemical} to arbitrary higher derivative theories and makes clear that there can be no contributions from boundary terms.

\label{horizon} 
\subsection{Near horizon behaviour}
Our results show that the transport coefficient $\xi$ is completely determined by the horizon behaviour of the bulk field $\phi(t,z)$. 
Following \cite{LiuMembrane}, we notice that infalling observers must see a regular field $\phi(t,z)$ at the horizon. This means that for these observers, $\phi$ can only depend on the ingoing Eddington-Finkelstein coordinate $v$ defined by
\be
dv=dt+\sqrt{-\frac{g_{zz}}{g_{tt}}}dz.
\ee
Actually time reversal symmetry implies that another solution is possible. Namely, it could also be the case that $\phi(t,z)$ depends on the outgoing Eddington-Finkelstein coordinate $u$,
\be
du=dt-\sqrt{-\frac{g_{zz}}{g_{tt}}}dz.
\ee
This means that
\bea
\partial_z \phi_0=\pm \sqrt{-\frac{g_{zz}}{g_{tt}}} \partial_t \phi_0=\mp  \frac{i \omega}{4\pi T}\frac {\phi_0}z \label{inorout}.
\eea
By composing the two possible behaviours (\ref{inorout}), it must be that the equation of motion at the horizon is given by
\be
\phi_{k}''(z)+\frac{\phi_{k}'(z)}{z}+\frac{\omega^2}{(4\pi T)^2}\frac{\phi_{k}(z)}{z^2}=0. \label{univeq}
\ee
Indeed this is what follows from the action \reef{lowaction} once we take the $z\to 0$ limit.
This argument is completely independent of the specific action one is considering, and so any action for the perturbation should lead to the same horizon behaviour. Regularity at the horizon is the only constraint. If we further demand on physical grounds that the perturbation should actually be infalling, the unique solution to the above equation is given by
\be
\phi_\omega(z)=\phi_0 \exp \left(-i\frac{\omega}{4\pi T} \log z\right). \label{nhsol}
\ee
This completely fixes the near horizon behaviour of the perturbation.

We see then that quite generally, close to the horizon the scalar perturbation must satisfy the universal equation \reef{univeq} regardless of the theory one is considering. This seems to be related to the conformal symmetry associated to the black hole horizon. If one looks at the form of the action discussed in subsection 3.2 one would have no reason to believe that the implied equation of motion would reduce to (\ref{univeq}) in the near-horizon limit. However, one must remember that the various functions involved are not arbitrary, and are only effectively deduced from an original scalar lagrangean density, which is necessarily smooth at the horizon.

Regularity then imposes that close to $z=0$ the effective action must be of the form
\bea
S_{\phi}^{(2)}&=& \int \prod_{i=1}^{d-1} dx^i \int \frac{d\omega}{2\pi} \int_{0}^1 dz \,\frac{ -\sqrt{-g}}{2\tilde \kappa} \left(g^{zz}\phi'_\omega(z)\phi'_{-\omega}(z)+g^{tt}\omega^2 \phi_{\omega}(z)\phi_{-\omega}(z)\right)\label{horac},
\eea
plus boundary terms. It is important to note that this form of the action does not mean that higher derivative terms simply drop out at the horizon. Indeed these play a crucial role in determining the effective horizon coupling constant $\tilde \kappa$. For obtaining this two derivative action from the general higher derivative one could follow the approach of \cite{Banerjee}. For instance, for a single higher derivative term of coefficient $\gamma$, one applies the lowest order equation of motion on the $\gamma$ terms, reducing them to terms with single derivatives or less. In this way one obtains a second order equation which can be deduced from an action of the form above in the near horizon limit. The problem with this procedure is that the overall coefficient of the action has to be fixed by hand by comparison to the original action\footnote{The authors of \cite{Banerjee} do this for higher derivative theories containing curvatures, but not their derivatives.}.

We will see that in practice one doesn't need to work with the equations of motion at all. It suffices to know that at the horizon the action must necessarily take the form \reef{horac}. For such an action, the canonical momentum associated is easily computed. At the horizon, using \reef{nhsol} we obtain\footnote{Note that at the horizon we have $\sqrt{-g}g^{zz} \simeq V (4\pi T)$.}:
\be
\Pi_{\omega}(z)=i \omega \frac{V}{\tilde \kappa}  \frac{\phi_0}z.
\ee
Using \reef{xifrompi} this gives
\be
\xi= \frac{V}{\tilde \kappa}. \label{xifromkappa}
\ee
In this way we have shown precisely the direct link between the transport coefficient $\xi$ and the effective coupling at the horizon $\tilde \kappa$, and how this is valid even for general higher derivative theories. 

\section{Membrane coupling as the pole at the horizon}
By now it is clear that we are only interested in the behaviour of quantities at the horizon. 
In this section we show that there is an efficient way of directly obtaining the value of the canonical momentum at the horizon, and therefore the membrane or horizon coupling $\tilde \kappa$. This is done by evaluating the lagrangian off-shell in a specific fashion. We then show how to apply this method to compute the shear viscosity and conductivity in higher derivative theories.

\subsection{The pole method}
Let us start by considering the near horizon form of the action, equation \reef{horac}.
Consider plugging into the action a perturbation of the form
\be
\phi_\omega(z)=\phi_0 \exp \left(-i\alpha \log z\right). \label{nhans}
\ee
In the near horizon limit we obtain
\be
S_{\phi}^{(2)}=\int \prod_{i=1}^{d-1} dx^i \int \frac{d\omega}{2\pi} \int dz \,\frac {V}{2\tilde \kappa} \left(\frac{\omega^2}{(4\pi T)^2}-\alpha^2\right) \frac{4\pi T}{z}\phi_0^2.
\ee
It is clear that on-shell the lagrangian is regular at the horizon, and indeed it is zero.
This is because it reduces to the boundary term $\partial_z(\Pi_\omega(z) \phi_{-\omega}(z))$, and the quantity being differentiated is a constant in the near horizon limit. However, we see that as long as we keep the perturbation off-shell, the lagrangian has a simple pole, the residue of which corresponds to the desired membrane coupling $\tilde \kappa$. In particular, starting from the full action, we might have considered a perturbation that only depended on $z$, effectively setting to zero the $\omega^2$ term in the above; or alternatively, a purely time dependent perturbation, which would set the $\alpha^2$ term above to zero. Either way, we can find out $\tilde \kappa$ from the residue at the pole, and use that information to get $\xi$ via \reef{xifromkappa}. 
To be precise, let us start off with an effective action for $\phi$ of the form:
\be
S^{(2)}_\phi=\int d^{d}x\, dz\, \mathcal L^{(2)}_\phi(\partial_z \phi, \partial_t \phi).
\ee
Once again our only requirement is that the lagrangian above does not depend on $\phi$ itself, but only on its derivatives, of which there can be an arbitrary number. Then we reach the following simple formulae for the transport coefficient $\xi$:
\bea
\xi &=& 8\pi T \lim_{\omega\to 0} \frac{\mbox{Res}_{z=0}\,\mathcal L^{(2)}_{\phi=z^{i \omega/(4\pi T)}}}{\omega^2} \qquad \mbox{Radial formula}\label{PoleRadialFormula} \\
\xi &=& -8 \pi T \lim_{\omega\to 0} \frac{\mbox{Res}_{z=0}\, \mathcal L^{(2)}_{\phi=e^{-i\omega t}}}{\omega^2} \qquad \mbox{Time formula}\label{PoleTimeFormula}.
\eea
We call this the pole method for deriving $\xi$. There is in fact an infinity family of formulae all exploiting the simple pole, but the above are the two simplest. 

This method works because essentially, the canonical momentum term in the lagrangian is the only one which develops a simple pole. In particular, all boundary terms have higher order divergences. This follows from the discussion of boundary terms in subsection \reef{gencanmom}. Since the action only contains differentiated fields, so do the boundary terms. Necessarily these will then be $\mathcal O(1/z^2)$ at the horizon.

On the other hand, the canonical momentum term necessarily has a simple pole, since $\Pi_\omega(z)=\Pi_\omega(0)$ in the low frequency limit. Using \reef{inorout} we obtain
\be
\lim_{z\to 0} \Pi_\omega(z) \phi_{-\omega}'(z)=-\frac{i\omega}{4\pi T} \frac{\Pi_{\omega}(0) \phi_0}z.
\ee
We see that the residue of the simple pole yields the value of the canonical momentum, which ultimately leads to $\xi$. Let us stress that the formulae \reef{PoleRadialFormula},\reef{PoleTimeFormula} are generic, and apply to any higher derivative lagrangian. In applications, this means for instance that we may compute the shear viscosity in theories with an arbitrary number of derivatives of curvatures. 

The only ingredients that go into these formulae are horizon regularity, and the requirement of a massless field. While the first is generically satisfied as long as we start from a covariant lagrangian, the second requirement is more restrictive. Notice also that knowledge of boundary terms is completely irrelevant, as long as the action only contains derivatives of the field. This can always be achieved by doing some by parts integration on the original action. In practice it is worth choosing our perturbations in a way that they can only appear differentiated, so as to avoid this extra work. We will see this is easily done for the shear viscosity and conductivity in the next section.

\subsection{Application to shear viscosity.}

The shear viscosity is computed on the field theory side by computing a specific two point function of the energy momentum tensor. More concretely, defining the retarded propagator
\be
G_R^{xy,xy}(\omega)=-i \int dt\,  \theta(t) \langle  T^{xy}(t)T^{xy}(0) \rangle e^{-i \omega t},
\ee
the shear viscosity is given by the Kubo formula
\be
\eta=\lim_{\omega \to 0} \frac{\mbox{Im}\,  G_R^{xy,xy}(\omega)}{i\omega}.
\ee
To compute the Green's function holographically we need to turn on a graviton perturbation $h_{xy}$. We focus on the shear mode channel at zero momentum, corresponding to a change of basis:
\be
dx_2\to dx_2+A_m(x^m) dx^m,
\ee
where $m$ runs through every coordinate except $x_2$.
In radial gauge and at zero momentum, the $A_{x_1}(t,z)\equiv \phi(t,z)$ component decouples from all others, so here we simply set them to zero.
It is clear from gauge invariance that $\phi(z)$ can never appear undifferentiated in the lagrangian, and so our formulae can be used. In particular, we need not worry about boundary terms.

Let us see how to use the pole method to compute the shear viscosity. Start by considering the following action
\be
\mathcal S=-\frac 1{16 \pi G_N}\int d^{5}x\sqrt{-g}\left(R+\frac{12}{L^2}+\gamma L^4 \nabla_{a}R_{bcde}\nabla^{a}R^{bcde}\right) \label{tryaction}
\ee
where $\gamma\ll 1$. The AdS-Schwarzschild metric,
\be
ds^2=\frac{L^2 dz^2}{4 z(1-z)^2(2-z)}+\frac{r_0^2}{L^2(1-z)}\left(-z(2-z)dt^2+\sum_i(dx_i)^2\right) \label{AdSBH}
\ee
with $i=1,..,3$ extremizes the action when $\gamma=0$. The associated temperature is easily found to be $T=r_0/(L^2\pi)$.
In the presence of the higher derivative term this is no longer a solution and receives an order $\gamma$ correction.

However, one does not need to compute the order $\gamma$ correction to the background metric if one is only interested in the $\eta/s$ ratio. This is a point which although known by experts, has not been made clear in the literature. If one is working to linear order in $\gamma$, then clearly it is sufficient to evaluate the higher derivative terms to the lagrangian on the lowest order background. On the other hand, in Einstein theory universality \cite{KSS} guarantees that $\eta/s=1/4\pi$ regardless of the details of the background we're considering.

Perturbing the metric by $A_{x_1}(t,z)=\phi(z)$ , and evaluating the lagragian to quadratic order gives:
\bea
S^{(2)}_{\phi}&=&-\frac{1}{32\pi G_N}\left(A\php'\phm'+B \php' \phm''+C\php''\phm''\right.\nonumber \\
&+& \left. D \php^{(3)}\phm'+E\php^{(3)}\phm''+F\php^{(3)}\phm^{(3)}\right) \label{eqnfunctions}
\eea
where the functions $A,B,...,F$ are given in appendix B. Let us start by computing the shear viscosity using the canonical momentum method. Applying definition \reef{canmom} gives
\be
\Pi_\omega(z)=\tilde A \php'(z)-(\tilde B \php'(z))'+(E \php''(z))''
\ee
with 
\bea
\tilde A&=& A-\frac 12 B'+\frac 12 D'' \nonumber \\
\tilde B&=&C-\frac 12 E'-D \nonumber.
\eea
Plugging in the near horizon solution \reef{nhsol} gives
\be
\Pi_\omega(0)=i \omega \frac{r_0^3}{16 \pi L^3 G_N}(1-1024 \gamma)\phi_0+\mathcal O(\omega^2),
\ee
from which we can read off the shear viscosity,
\be
\eta=\frac{1}{16 \pi G_N}\left(\frac{r_0^3}{L^3}\right)(1-1024 \gamma).
\ee
Now, let us try the pole method. We simply evaluate the lagrangian density associated to the action \reef{tryaction} on the perturbed background metric to quadratic order in the field $\phi$.
Substituting $\phi(z)=z^{i \omega L^2/(4 r_0)}$ and taking $z\to 0$ we obtain
\be
\mathcal L=\frac{1}{16\pi G_N}\left(...+ \frac{(\omega r_0)^2}{8 L} \frac{1-1024 \gamma}{z}+ \mbox{Regular}\right)
\ee
Had we evaluated the lagrangian on a metric perturbed by $A_{x_1}(t,z)\equiv \phi(t)=e^{-i \omega t}$ we would get the same result but with an opposite sign. From this expression we read off exactly the same value for $\eta$ as by the canonical momentum method. Computationally this is highly efficient: one simply evaluates the full higher derivative lagrangian on a perturbed metric and looks for the pole at $z=0$. 

\subsection{Application to DC conductivity}

The computation of the DC conductivity can also be achieved using our method. 
Starting from the retarded propagator,
\be
G_R^{x,x}(\omega)=-i e^2 \int dt\,  \theta(t) \langle  J^{x}(t)J^{x}(0) \rangle e^{-i \omega t},
\ee
the conductivity is given by the Kubo formula
\be
\sigma=\lim_{\omega \to 0} \frac{\mbox{Im}\,  G_R^{x,x}(\omega)}{i\omega}.
\ee
Here $e$ is the coupling of the weakly gauged boundary field theory \cite{KovtunRitz}.

Consider a general action of the form
\be
S=-\frac{1}{2 g_{d+1}^2}\int d^{d}x\, dz \sqrt{-g}\left(\frac 1{4} X^{a b c d}F_{ab}F_{cd} \right).
\ee
We can equivalently define the tensor $X^{abcd}$ by
\be
X^{abcd}=-\frac{4 g_{d+1}^2}{\sqrt{-g}} \frac{\delta \mathcal L}{\delta F_{ab} \delta F_{cd}}.
\ee
We assume there is no background gauge field present. Turning on a small perturbation $A_{x_1}(t,z)\equiv \phi(t,z)$, the resulting quadratic effective action does not involve undifferentiated perturbations:
\be
S^{(2)}_{\phi}=-\frac{1}{2 g_{d+1}^2}\int d^{d}x\, dz\, \sqrt{-g}\,g^{xx}\,\left(X^{z x_1}_{~ ~ ~ z x_1}g^{zz} \partial_z\phi \, \partial_z \phi+
X^{t x_1}_{~ ~ ~ t x_1}g^{tt}\, \partial_t\phi\, \partial_t \phi\right)
\ee
Applying the formulae \reef{RadialFormula},\reef{TimeFormula} it is straightforward to find:
\be
\sigma=\left(\frac{e^2}{g_{d+1}^2}\,g_{xx}^{d-3}\, X^{z x_1}_{~ ~ ~ z x_1}\right)\Bigg |_{z=0}= \left(\frac{e^2}{g_{d+1}^2}\,g_{xx}^{d-3}\, X^{t x_1}_{~ ~ ~ t x_1}\right)\Bigg |_{z=0}. \label{sigma2Der},
\ee
where the last equality is required by regularity of the effective action at the horizon.
Now consider the special case
\be
X_{abcd}=g_{ac}g_{bd}-g_{bc}g_{ad}-8\gamma \,C_{abcd}
\ee
where $C_{abcd}$ is the Weyl tensor. For the AdS-Schwarzschild metric in $d+1$-dimensions we obtain 
\be
C^{t x_1}_{~ ~ ~ t x_1}=C^{z x_1}_{~ ~ ~ z x_1}=-\frac{(1-z)^2}{L^2}\to -\frac{1}{L^2}
\ee
and therefore
\be
\sigma=\frac{e^2}{g^2}\left(\frac{r_0}{L}\right)^{d-3}\left(1+\frac{8\gamma}{L^2}\right)
\ee
in agreement with \cite{KovtunRitz},\cite{Ritz}.

\section{Wald-like formulae for transport coefficients}

From our results in the previous sections, it is clear that there exists a class of transport coefficients which are holographically given by effective couplings at the horizon, or membrane couplings. There is a specific proposal in the literature for a formula that gives this coupling for the special case of the shear viscosity \cite{BrusteinViscosity}. It has been shown that this formula is incomplete as it does not yield the correct answer for well known case of a $C^4$ correction \cite{Banerjee}. It is therefore of interest to find such a formula, which we do so in this section for a large class of lagrangians. We also give an analogous expression for DC conductivity. 
Equations \reef{RadialFormula}, \reef{TimeFormula} and \reef{CondForm} are the main results of this section. 

\subsection{Deduction of formulae}
We start in the background \reef{bg} and add a shear mode perturbation
\be
dx_2\to dx_2+A_m(x^n) dx^m,
\ee
as before. The indices $m,n,p,...$ do not include $x_2$. Under such an addition the curvature transforms as \cite{MyersShenker}
\bea
R_{mnpq}&=& \hat R_{mnpq}-\frac 34 e^{2\rho} P[F_{m n}F_{p q}]\nonumber \\
R_{m y m y}&=& \hat R_{m y m y}+\frac 14 e^{4\rho} F_{m p}F_{n}^{\, p} \nonumber \\
R_{mnp y}&=&-\frac 12 e^{-\rho} P[\nabla_p(e^{3\rho} F_{m m})]
\eea
where hatted quantities refer to the unperturbed background and where $P$ acting on a tensor gives it the relevant symmetries for the Riemann tensor. For instance
\be
P[\nabla_p (e^{3\rho} F_{mn})]=\frac 13\left[2 \nabla_p(e^{3\rho} F_{mn})+\nabla_n(e^{3\rho} F_{mp})-\nabla_m(e^{3\rho} F_{np})\right]
\ee
We consider a general lagrangian which does not involve derivatives of curvatures. We assume an action of the form
\be
S=-\frac 1{16\pi G_N} \int d^d x \sqrt{-g} \, \mathcal L(R_{abcd},\tilde F_{ab}^{(q)},\Phi^{(r)},...)
\ee
The action may depend on an arbitrary number of other fields such as abelian gauge fieds and scalars. All we require is that the shear mode appears only through the curvature, which in practice requires that the field strengths do not appear differentiated. Scalar fields can have single derivatives on them, or an arbitrary number of box operators.

Using the expressions above, the quadratic effective action $S^{(2)}$ for the shear mode is given by
\bea
S^{(2)}&=&-\frac{1}{32\pi G_N}\int dy \int d^d x\sqrt{-g} \left\{\left[-\frac 34 e^{2\rho}X^{mnpq} F_{mn}F_{pq}+n_{a}n_{b}\, X^{a m b n} e^{2\rho} F_{m p}F_{n}^{\, p}\right] \right. \nonumber \\
&& \qquad + 2 \, Y^{a mnp, b qrs}\Bigg [n_{a}n_{b}\,e^{-4\rho} \nabla_m(e^{3\rho} F_{np})\nabla_q(e^{3\rho} F_{rs})\Bigg ]\Bigg \} \label{shearac}
\eea
where
\be
X^{abcd}=-\frac{8 \pi G_N}{\sqrt{-g}}\frac{\delta \mathcal L}{\delta R_{a b c d}}, 
\qquad Y^{abcd,efgh}=\frac{\delta X^{abcd}}{\delta R_{efgh}}
\ee
and $n_{x_2}=e^{\rho}$, all other components zero. Notice that $Y,X$ inherit all the symmetries of the Riemann tensor.

Taking $A_{x_2}(t,z)\equiv \phi(z)=z^{-i \omega/(4\pi T)}$ and expanding the effective lagrangian near the horizon, we can read off the residue at the pole and obtain the shear viscosity. Our calculations are given in detail in appendix A. The final result is
\be
\eta=\frac{V}{16 \pi G_N}\Bigg(2 X^{zy}_{~~zy}-X^{xy}_{~~xy}-\frac{4\, e^{-2g_0}}{L^2}\,\Bigg[\partial_z \alpha^{zt}-\alpha^{zt}\left(e^{2g_0}L^2 R+(d+3)\partial_z \rho\right)\Bigg]\Bigg)\Bigg |_{z=0} \label{RadialFormula}
\ee
where 
$$
\alpha^{zt}=Y^{xz~~,yz}_{~~~zy~~~zx}-Y^{xz~~,yt}_{~~~zy~~~tx} \label{alphas}
$$
and we have renamed $x_1\equiv x, x_2 \equiv y$.

Alternatively one may take $A_{y}(t,z)\equiv \phi(t)=\phi_0 e^{-i \omega t}$ and reading off the residue at the pole get the formula,
\be
\eta=\frac{V}{16 \pi G_N}\Bigg(2 X^{ty}_{~~ty}-X^{xy}_{~~xy}-\frac{4\, e^{-2g_0}}{L^2}\Bigg[\partial_z \tilde \alpha^{zt}-\tilde \alpha^{zt}\left(e^{2g_0}L^2 R+(d+3)\partial_z \rho\right)\Bigg]\Bigg)\Bigg|_{z=0} \label{TimeFormula}
\ee
where
$$
\tilde \alpha^{zt}=Y^{xt~~~,yz}_{~~~zy~~~tx}+Y^{xt~~\,,y~z}_{~~~zy~\,t~x}.
$$
These formulae are similar to Wald's formula for the entropy \cite{Wald}. In particular, in the background \reef{bg} and in our notation the entropy density is given by
\be
s=\frac V{4\, G_N}\, X^{zt}_{\ \ zt}\, \Bigg |_{z=0}
\ee
Notice however that in this case the double Riemann derivative type coefficients do not contribute. This is because the viscosity is related to metric fluctuations which involve differentiating the action twice with respect to the metric.

We can now do the same sort of calculation for the conductivity. We consider an action for the gauge field of the form
\be
S_A=-\frac 1{2g_{d+1}^2}\int d^{d+1}x \sqrt{-g}\left(-\frac 14 X^{abcd}F_{ab}F_{cd}+Y^{abc,def}\nabla_a F_{bc}\nabla_d F_{ef}\right),
\ee
and once again assume there is no background gauge field turned on. Then it is straightforward to generalize the result \reef{sigma2Der}:
\be
\sigma=\frac{e^2}{g_{d+1}^2}\,g_{xx}^{d-3}\Bigg( X^{z x_1}_{~ ~ ~ z x_1}-\frac{2\, e^{-2g_0}}{L^2}\Bigg[ \partial_z \beta^{zt}-\beta^{zt}\left (e^{2g_0}L^2 R+(d-1) \partial_z \rho\right)\Bigg]\Bigg)\Bigg |_{z=0}, \label{CondForm}
\ee
with
$$
\beta^{zt}=Y^{z~~,z}_{~zy~~zy}-Y^{z~~,t}_{\ zy~~ty}.
$$
Just like for the shear viscosity there is also another formula obtained by taking a time dependent perturbation.
We have tested these formulae against the generalized canonical momentum and pole methods and have found agreement. 

\subsection{Comparison with previous proposal}

The formulae deduced in the previous sections are perturbatively valid in the coefficients controlling the size of higher derivative corrections, but they are correct to any desired order one wishes to consider. In theories containing only two derivatives, such as Lovelock gravity, the formulae are non-perturbatively correct.

Our formulae for the shear viscosity are distinctively different from the proposal of \cite{BrusteinViscosity}. For backgrounds of the form \reef{bg} and in our notation the proposal is
\be
\eta=\frac{V}{16 \pi G_N}\, X^{xy}_{~ ~ xy}\Bigg |_{z=0}. \label{brustein}
\ee
The authors of \cite{BrusteinViscosity} found that this formula gives the correct shear viscosity for Gauss-Bonnet theory. Let us see precisely why this happens. Consider then Gauss-Bonnet theory in five dimensions. The action is given by
\be
S=-\frac 1{16\pi G_N}\int d^{5}x \sqrt{-g}\left(R+\frac{12}{L^2}+W\right) \label{higherderac}
\ee
with correction $W=\frac{\lambda}2 L^2\left(R_{abcd}^2-4 R_{ab}^2+R^2\right )$. The action is extremized by the black hole solution
\be
ds^2=\frac{r_0^2}{L^2}\left(-N^2 \,f(u)dt^2+dx^i dx_i\right)+\frac{du^2}{4\, u^2 f(u)}
\ee
with
\be
f(u)=\frac 1{2\lambda}\left(1-\sqrt{1-4 \lambda\,(1-u^2)}\, \right), \qquad N^2=\frac 12 \left(1+\sqrt{1-4 \lambda}\right).
\ee
This background can be put in the general form \reef{bg} by doing the substitution $u\to 1-z$.
Computing the various quantities in the formula \reef{RadialFormula} we obtain
\bea
X^{xy}_{~ ~xy}\Bigg |_{z=0}&=& 1-4\lambda-32\lambda^2, \qquad X^{zy}_{~ ~ zy}\Bigg |_{z=0}=1-8\lambda \nonumber \\
\partial_{z}\alpha^{zt}\Bigg |_{z=0}&=& 0, \qquad \alpha^{zt}\Bigg |_{z=0}=\frac{\lambda L^2}4, \qquad R~ \Bigg|_{z=0}=-\frac{20}{L^2}\left(1+\frac 85 \lambda\right)
\eea
It is immediately apparent that the formula of \cite{BrusteinViscosity} cannot be correct, as it leads to quadratic terms in $\lambda$. We also see that by an apparent accident, the formula gives the right result perturbatively in $\lambda$. Plugging in these values into \reef{RadialFormula}, quadratic terms cancel and we get
\be
\eta=\frac{1}{16 \pi G_N}\left(\frac{r_0}{L}\right)^3(1-4 \lambda),
\ee
which indeed is the correct value in five dimensions \cite{KatsPetrov}\cite{MyersShenker}. To obtain this result it was crucial to include the double-Riemann derivative type coefficients.

Now consider applying the formulae for the well known $C^4$ corrections. In this case
\be
W=\gamma L^{6}\left(C_{a b c d} C^{a b}_{\ \ e f} C^{c \ d}_{\ g \ h} C^{e g
f h}-\frac 12 C_{a b c d} C^{a b}_{\ \ e f} C^{c e}_{\ \ g h} C^{d g
f h}\right ).
\ee
The computation of the various coefficients involved is done straightforwardly with the aid of a computer\footnote{Notebooks are available from the author upon request.}. We obtain
\bea
X^{zy}_{~ ~ zy}\Bigg |_{z=0}&=&1-20\gamma, \qquad X^{xy}_{~ ~xy}\Bigg |_{z=0}= 1+20\gamma  \nonumber \\
\partial_{z}\alpha^{zt}\Bigg |_{z=0}&=&-6\,\gamma L^2 , \qquad \alpha^{zt}\Bigg |_{z=0}=\frac 32\, \gamma L^2.
\eea
Naively applying formula \reef{brustein} gives
\be
\eta=\frac{1}{16 \pi G_N}\left(\frac{r_0}{L}\right)^3(1+20\gamma).
\ee
This result had already been reported in \cite{Banerjee}. Our formula however gives
\be
\eta=\frac{1}{16 \pi G_N}\left(\frac{r_0}{L}\right)^3(1+180\gamma).
\ee
which matches the previous calculation in the literature \cite{BuchelViscosity}\cite{Disagreement}.

This disagreement can be ultimately traced to an incorrect formula for the kinetic term for gravitons. It seems that this is not limited to the $h_x^y$ gravitons, and so in the same fashion one expects that it is not correct to think of the Noether charge entropy as the effective coupling at the horizon of $h_{rt}$ gravitons \cite{BrusteinEntropy}. Our results show that even in two-derivative theories the graviton kinetic term receives contributions double-Riemann derivative type terms.
\section{Extremal backgrounds}

In this section we will be interested in studying the holographic duals of field theories at zero temperature but finite chemical potential.
In this regime, the holographic dual gravitational background is expected to contain an extremal black hole. It is then possible to take a scaling limit where one focuses on the near horizon geometry \cite{LiuAdS2}. We will consider theories where this near horizon geometry is given by $AdS_2\times R^{d-1}$, supported by the flux of an abelian gauge field $F_{ab}$. In terms of the dual field theory, the full geometry describes an RG flow between the boundary $d$-dimensional field theory and a conformal fixed point located in the IR, characterized by the $AdS_2$ factor. The near horizon geometry is described by the following metric and gauge field:
\bea
ds^2&=&-v_1\left(-z^2 d\tau^2+\frac{dz^2}{z^2}\right)+v_2(dx^2). \nonumber \\
F_{z\tau}&=& Q.
\eea
The equations of motion fix $Q$ in terms of the $AdS_2$ radius $v_1$. In this background, a massless scalar field $\phi(z,\tau)=\phi(z)e^{i\omega \tau}$ satisfies the equation
\be
\phi''(z)+\frac{2}{z^2}\phi'(z)+\frac{\omega^2}{z^4}\phi(z)=0. \label{nhExt}
\ee
This is exactly the equation of motion one would expect close to a double pole horizon. Indeed for a metric of the form,
\bea
ds^2=\frac {L^2}{z^2}\, e^{2g(z)} dz^2+\left(-z^2\, e^{2f(z)} dt^2+e^{2\rho(z)} dx^i dx_i\right) \label{bg2},
\eea
regularity in Eddington-Finkelstein coordinates at the horizon imposes
\bea
\partial_z \phi_0=\pm \sqrt{-\frac{g_{zz}}{g_{tt}}} \partial_t \phi_0=\mp  \frac{i\omega}{\mu}\frac {\phi_0}{z^2} \label{inorout2},
\eea
where we have defined the chemical potential $\mu=e^{f_0-g_0}/L$. By composing the two possible behaviours we get back \reef{nhExt} after a suitable redefinition of $\omega$. The solution to \reef{nhExt} is given by
\be
\phi(z)=\phi_0\exp\left(\pm \frac {i \omega}z\right).
\ee
From the point of view of the full geometry, this only fixes the near horizon behaviour of the perturbation. But this is all that is necessary to calculate the canonical momentum, as we've seen in the previous sections. For a two derivative theory
\be
\pi_\omega=-\frac 1{\kappa}\sqrt{-g}g^{zz} \partial_z \phi_\omega(z)=i (\omega \phi_0)\,\frac{V}{16 \pi G_N} .
\ee
The existence of a double pole hasn't changed anything. The canonical momentum is still regular at the horizon, and from it we can read off the transport coefficient associated with $\phi$, $\xi=V/\kappa$. This immediately implies the universality of $\eta/s=1/4\pi$ in these theories. 

It is clear that this type reasoning does not change when we go to higher derivative theories, and the generalized canonical momentum method described in section 3 is still valid for these backgrounds.

\subsection{Shear viscosity}
It seems that in these extremal backgrounds the analytic formula for the shear viscosity and the pole method have to be modified. This is because their deduction crucially relied on the presence of a simple pole in the action, which is no longer the case here, since the $\Pi(z)\phi'(z)$ term is now $\mathcal O(1/z^2)$. However the structure of the canonical momentum as a function of $\phi'(z)$ is necessarily the same at zero and finite temperature. That is, at the horizon the canonical momentum is always of the form %
\be
\Pi(z)=-\frac {\sqrt{-g}}{\tilde \kappa}  g^{zz} \partial_z \phi(z)= \frac {i \omega}{\tilde \kappa} \sqrt{-g} g^{zz} \sqrt{-\frac{g_{zz}}{g_{tt}}}\phi_0.
\ee
Whether $g_{zz}$ has a simple or a double pole is irrelevant in the above calculation, since in the end we are only interested in finding the coefficient $\tilde \kappa$. This means that the passage from finite to zero temperature is smooth, and so we are free to compute the shear viscosity at finite temperature and then take the zero temperature limit. This will necessarily yield the correct strictly zero temperature result. Therefore we can simply take the background off extremality, do the computation there and then extrapolate to $T=0$.

While this suffices to make the analytic formula and the pole method work, we can actually do better, since we are not fully exploring the symmetries of the problem in this fashion. Since we are only interested in the near-horizon behaviour, we can directly use the radial formula for shear visocisty in the $AdS_2\times R^{d-1}$ geometry. It is simple to see that this leads to
\be
\eta=\frac{V}{16 \pi G_N}\left(2 X^{zy}_{~~zy}-X^{xy}_{~~xy}-8\frac{\alpha^{zt}}{v_1} \right). \label{etaX}
\ee
The coefficient $\partial_z \alpha^{zt}$ has vanished as there can be no $z$ dependence by symmetry.

It should be emphasized that this formula is only valid assuming that there are no covariant derivatives of the gauge field strength in the action; otherwise the shear viscosity receives extra contributions from this type of terms. As for the pole method, it can also be directly applied in the near horizon geometry, and it works for any higher derivative lagrangian. We do need to work at finite temperature in order to have simple poles. This is accomplished with the $AdS_2$ black hole geometry,
\bea
ds^2&=&-v_1\left(-(z^2-z_0^2)  d\tau^2+\frac{dz^2}{z^2-z_0^2}\right)+v_2(dx_i dx^i),\\
F_{z\tau}&=& Q,  \label{extBg}
\eea
which is obtained by taking the $T\to 0$ and near horizon limit simultaneously as in \cite{LiuAdS2}.

Higher derivative corrections modify $v_1,v_2$ but leave the form of the metric invariant. Then the residue formulas are directly appliable, and the final result does not depend on the parameter $z_0$, as expected.

Let us see how this works in practice. Consider the action
\be
S=-\frac 1{16 \pi G_N} \int d^{5}x\sqrt{-g} \left(R+\frac{12}{L^2}-\frac 1{4e^2} F_{ab}F^{ab}+\gamma L^2 R_{abcd}R^{abcd}\right) \label{SExt}
\ee
with $\gamma\ll 1$. This action has a solution of the type $AdS_2 \times R^{3}$, corresponding to the near horizon limit of the extremal AdS-Reissner-Nordstrom black hole. The parameters $v_1, v_2, Q$ are given by
\be
v_1=\frac{L^2}{12}, \qquad v_2=\frac{r_0^2}{L^2}, \qquad Q=\frac{L^2}{\sqrt{6}}(1-24\gamma) \label{solExtBg}
\ee
The temperature is then given by $T=z_0/(2\pi)$.
Adding a shear mode perturbation of the form $\phi(z)=e^{-i \omega t}$ leads to a pole of the lagrangian at $z=z_0$. The residue is
\be
\mbox{Res}_{z=0}\mathcal L^{(2)}_{\phi=e^{-i\omega \tau}}=-\frac{\omega^2}{4\, z_0} \frac{1}{16 \pi G_N} \left(\frac{r_0}{L}\right)^3(1-48 \gamma).
\ee
Applying the time formula \reef{PoleTimeFormula} gives
\be
\eta=\frac{1}{16 \pi G_N} \left(\frac{r_0}{L}\right)^3(1-48 \gamma).
\ee
This result agrees with the one obtained in \cite{Chemical}. Alternatively, one may use the formula \reef{etaX}. The various quantities are determined to be
\bea
2X^{zy}_{~~zy}-X^{xy}_{~~xy}=1, \qquad \alpha^{zt}=\frac{ \gamma L^2}2,
\eea
leading to the same result. We have tested both these methods for various corrections by comparing them with the generalized canonical momentum method, finding always an agreement.

\subsection{Conductivity}

Now let us consider the conductivity. Here the story is more interesting due to the presence of a background gauge field. Because of this the computation of the conductivity is no longer determined solely by the horizon behaviour. The background charge acts as an effective mass for the gauge field perturbation, leading to a non-trivial flow of the associated canonical momentum from the horizon to the boundary.

In the extremal $T=0$ limit, we will still able to unambiguously determine the low frequency limit of the conductivity. This is because the frequency dependence of the imaginary part of the Green's function is completely fixed by the near horizon behaviour. The flow modifies the norm of the perturbation from the horizon to the boundary, but it cannot affect the frequency dependence. As shown in \cite{LiuAdS2},
\be
\mbox{Im} G_{R}(\omega)=G_R(\omega=0) d_0 \mbox Im \mathcal G(\omega) \propto \omega^{2 \nu_k} \label{flow}
\ee
where $d_0$ is some frequency independent flow factor, $\mathcal G(\omega)$ is the two point function computed in the IR CFT, and $\nu_k$ is related to the conformal dimension of the operator under question. Let us see how this works in practice. 

Consider the Einstein-Maxwell action with a cosmological constant in $d+1$-dimensions, and focus on the $AdS_2\times R^{d-1}$ solution supported by flux described in the previous section. Now turn on small perturbations corresponding to the shear channel:
\bea
a_{x}(t,z)&=& A(z)e^{-i\omega \tau} \nonumber \\
h_t^{x}(t,z)&=& H(z)e^{-i\omega \tau} \nonumber \\
h_r^{x}(t,z)&=& R(z)e^{-i\omega \tau} \nonumber \\
h_y^{x}(t,z)&=& \phi(z)e^{-i\omega \tau} \label{pert}
\eea
The $\phi(z)$ perturbation decouples from the others. Working in radial gauge where $R(z)=0$ we get two equations of motion plus a gauge constraint
\bea
A_x''(z)+\frac 2{z} A_x'(z)+\frac{\omega^2+z^2 v_2 H'(z)/v_1}{z^4} A_x(z)&=&0, \nonumber \\
Q A'(z)+v_2 H''(z)&=& 0, \nonumber \\
Q A(z)+v_2 H'(z)&=& 0. 
\eea
Solving the gauge constraint and plugging it into the first equation of motion we obtain
\be
A''(z)+\frac 2z A'(z)+\frac{\omega^2-z^2 Q^2/v_1}{z^4} A(z).
\ee
This is the equation of motion for a massive scalar field with $m^2=Q^2/(v_1 L^2)$. In the Einstein-Maxwell case this gives us $m^2=2/L^2$, corresponding to a conformal dimension $\delta=1/2+\nu=2$. This equation is easily solved, with solution
\be
A(z)=\exp(i\omega/z)(z-i\omega) 
\ee
where we have chosen infalling boundary conditions. Expanding around the boundary this gives
\be
A(z)\simeq z+\frac{\omega^2}{2z}+\frac 13 \frac{i \omega^3}{z^2}+...
\ee
This implies \cite{LiuAdS2}, \cite{Leigh} that the IR Green's function is given by
\be
\mathcal G_\omega=i \omega^3.
\ee
The flow equation \reef{flow} then fixes
\be
\mbox{Im} G_R(\omega)\propto i \omega^3 \Rightarrow \mbox{Re}(\sigma)\propto \omega^2.
\ee
This result is universal as long as the geometry contains a flux supported $AdS_2$ factor. Since we are focusing on the imaginary part of $G_R$ our calculations say nothing about the imaginary part of the conductivity. Indeed, generally there one expects a pole at $\omega=0$, which by Kramers-Kronig relations indicates a delta function for the DC conductivity\cite{Sachdev}.\footnote{We thank Sean Hartnoll for pointing this out. Throughout this section we focus on the low frequency limit of the real part of the conductivity.}

We have considered taking the zero temperature limit with $\omega/T\to \infty,\,\, \omega/\mu\to 0$. The analogous computation with
$\omega/T\to 0, \,\, \omega/\mu\to 0$ corresponds to taking the zero temperature limit of the finite temperature conductivity. In that case we also get a zero conductivity limit, except for the special case $d=2+1$, where it is a constant \cite{KovtunRitz}.

However, there is yet another way of taking the zero temperature limit, and that is by requiring $\omega/T\to z_0,\,  \omega/\mu\to 0$. In this limit we are working with the $AdS_2$ black hole background. \footnote{The other two limits then correspond to taking $z_0\to 0$ where we get back pure $AdS_2$, or $z_0\to +\infty$ where the $AdS_2$ space disappears and we are left with a regular non-extremal black hole.}

The equation of motion for the gauge perturbation is then
\be
A''(z)+\frac {2z}{z^2-z_0^2} A'(z)+\frac{\omega^2-(z^2-z_0^2)Q^2/v_1}{z^4} A(z).
\ee
The solution of this equation that is infalling at the horizon is
\be
A(z)=z_0\exp\left(-\frac{\pi \omega}{2z_0}\right) \Gamma\left(2-\frac{i\omega}{z_0}\right) \mathcal L_P\left [1,\frac{i\omega}{z_0}, \frac z{z_0}\right ]
\ee
with $\mathcal L_P$ an associated Legendre polynomial. This solution has the boundary expansion
\be
A(z)\simeq z+\frac{\omega^2}{2z}+\frac 38 \frac{i \omega^3}{z^2}
\ee
from which we read off
\be
\mathcal G_{R}(\omega)=\frac 98 i \omega^3.
\ee
Therefore in this case we still obtain zero conductivity in the low frequency limit.

\subsection{Conductivity with arbitrary higher derivative corrections}
Now let us see what happens when one includes higher derivative corrections to the Einstein-Maxwell action. It is instructive to consider the simple case where the action is defined by \reef{SExt}. We take the background geometry as in \reef{extBg},\reef{solExtBg}. The equations of motion for the perturbations \reef{pert} can be derived using an effective action approach (see e.g. \cite{Benincasa}). Here we show the gauge perturbation equation together with the gauge constraint:
\bea
A''(z)+\frac 2{z} A_x'(z)+\frac{\omega^2}z^4 A(z)+ 2 \sqrt{6} v_2 \frac{H'(z)}{z^2}&=& \gamma S_A \nonumber \\
\frac 1{\sqrt{6}}A(z)+v_2 H'(z)&=& \gamma S_B. \nonumber
\eea
The sources are
\bea
S_A&=& -48 \sqrt{6} \frac{H'(z)}{z^2} \nonumber \\
S_B&=& 4\frac{-\sqrt{6} z^2 A(z)+12 v_2\left[(\omega^2-3 z^2) H'(z)+z^3(2H''(z)+z H'''(z))\right]}{z^2}.
\eea
Working perturbatively in $\gamma$, we apply the lowest order equations of motion on the sources to simplify them. In the end we get two simpler equations:
\bea
A''(z)+\frac 2{z} A_x'(z)+\frac{\omega^2}z^4 A(z)+ 2 \sqrt{6} v_2 \frac{H'(z)}{z^2}+48 \gamma \frac{A(z)}{z^2}&=&0 \nonumber \\
\frac {1+24 \gamma}{\sqrt{6}}A(z)+v_2 H'(z)&=& 0 \nonumber.
\eea
Plugging the second equation into the first we get
\be
A''(z)+\frac 2z A'(z)+\frac{\omega^2-2 z^2}{z^4} A(z)=0. \label{gaugeeq}
\ee
Miraculously, the $\gamma$ dependence has completely dropped out. One could in principle expect that the ``mass term'' could receive $\gamma$ corrections. The fact that it doesn't means that the conformal dimension associated to $A(z)$ does not get renormalized by the curvature squared correction to the lagrangian.

This result seems to be completely general. We have repeated this procedure for a variety of higher derivative theories, including terms of type $RFF, (\nabla F)^2, (\nabla \nabla F)^2, R (\nabla F)^2, F^4$,etc. We have also done it in the $AdS_2$ black hole geometry, and have even performed these perturbative computations to second order\footnote{Notebooks are available from the author upon request.}. The equation of motion for $A(z)$ stubbornly remains the same, although the gauge constraint receives corrections at all orders.

Actually, this result also applies to the shear viscosity calculation. Higher derivative corrections to the equation of motion for $\phi(z)$ defined in \reef{pert} can also be treated perturbatively, and there one also gets back the original lowest order equation of motion. Of course there we have an understanding on why this is the case, since the equation of motion for a massless scalar is essentially fixed by demanding horizon regularity, or equivalently it is fixed by the conformal symmetry of the background. Even though the same argument necessarily fixes the form of the equation of motion for the gauge perturbation $A(z)$ deep inside the $AdS_2$ bulk, the situation is different because of the presence of a background charge. This acts as an effective mass, producing a flow from the $AdS_2$ horizon to the boundary. There seems to be in principle no reason why higher derivative corrections shouldn't modify this flow.

Even though we haven't been able to formally prove that the equation \reef{gaugeeq} is unaffected by perturbative corrections to the lagrangian, our results strongly indicate this is the case. We therefore conjecture that this is true to all orders in perturbation theory. This means that from the point of view of the IR CFT, the scalar operator $a_x(t,r)$ has protected conformal dimension $\delta=2$.

This result has important consequences for the conductivity of field theories at extremality. Since the frequency dependence of the Green's function doesn't change, we conclude that the real part of the conductivity scales like $\omega^2$ in the low frequency limit, to all orders in the higher derivative expansion.

\section{Discussion}

In this paper we have shown that the approach of \cite{LiuMembrane} is generalizable to higher derivative theories. Thermal retarded Green's functions are given by the generalized canonical momentum. In the hydrodynamic, low frequency limit where the flow of this quantity is trivial, regularity at the horizon completely determines its form up to a constant. Similarly, we've also seen how at the horizon the action is also specified up to that same constant factor, the effective horizon or membrane coupling, which essentially determines the associated transport coefficient. We would like to note however, that our method only applies for first order transport coefficients, that is, those that can be obtained by a Kubo formula in the low frequency limit. Higher order coefficients require going beyond the low frequency and zero momentum limit, and in this case one has a non-trivial flow from the horizon to the boundary which must be taken into account.

To determine the membrane coupling one exploits the form of the action at the horizon to generate poles in the lagrangian density. The residue of the simple pole then simply determines the associated transport coefficient. The pole method given by equations \reef{PoleRadialFormula} and \reef{PoleTimeFormula} is generally applicable to any higher derivative theory, and requires only the evaluation of the lagrangian on a perturbed background, making it extremely efficient. We hope this method will facilitate future computations of corrections to transport coefficients. 

We have also derived analytic formulae for the shear viscosity and conductivity, in equations \reef{RadialFormula},\reef{TimeFormula},\reef{CondForm}. Though these formulae are not fully covariant, it should be possible to make them so with some work. It is easily checked that they do respect radial coordinate reparameterizations. Our formula for the shear viscosity shows that the proposal of \cite{BrusteinViscosity} is incomplete, as it does not properly account for the contribution of higher derivative terms in the action.

Ultimately our methods work because of the specific constraints that arise at the horizon. This seems to be related to an emerging conformal symmetry which completely fixes the behaviour of any perturbations. It would be interesting to understand this in more detail.

We have shown that the pole method for computing the shear viscosity is still applicable in extremal black hole backgrounds. Alternatively one can use the simple formula \reef{etaX}. We have found that in the dual IR CFT there is an operator which seems to have protected conformal dimension to all orders in perturbation theory. It might be possible to obtain a proof of this statement, and understand generically which operators satisfy this property by performing a dimensional reduction onto an effective two dimensional theory. This would be important as in our case this fact is responsible for fixing the low frequency scaling of $\mbox Re(\sigma)$ to be $\mathcal O(\omega^2)$.

It seems likely that our methods can be extended more or less straightforwardly to backgrounds dual to field theories with non-relativistic symmetries \cite{hotspacetimes},\cite{Lifshitz}. This is because these backgrounds have similar horizon structure as in the relativistic case. Also it would be interesting to see if we can adapt our methods to study fermionic correlation functions. We leave these and other questions for future work.

\acknowledgments

It is a pleasure to acknowledge the useful comments and suggestions of Robert Myers, Aninda Sinha, and Alex Buchel. The author would also like to thank the Perimeter Institute for hospitality while part of this project was realized. This work was supported by the Portuguese government, FCT grant SFRH/BD/23438/2005.

\appendix
\section{Derivation of the formulae for the shear viscosity.}

We start with the geometry \reef{bg}. The Christoffel symbols are
\bea
\Gamma^{z}_{\ zz}&=&-\frac 1{2z}+g'(z)\qquad \Gamma^{t}_{\ tz}=\frac 1{2z}+f'(z)\qquad \Gamma^{x_i}_{\ x_iz}=\rho'(z).
\eea
The strategy is to find the simple pole contained in the effective action for the shear mode \reef{shearac}. These come from $\phi'^2$ factors upon plugging in the near horizon form of the perturbation:
\be
\sqrt{-g}g^{zz}\phi_\omega'(z) \phi_{-\omega}'(z)\to \frac{V}{z}\frac{\omega^2}{4\pi T}.
\ee
It is also useful to notice that at the horizon we have
\bea
\nabla_z F^{z}_{\ y}\simeq -\nabla_t F^{t}_{\ y}\mathcal O(1/z), \quad \nabla_{x_i} F^{x_i}_{\ y}\simeq \mathcal O(1) 
\eea

Start by considering the two derivative terms. We define $x_2\equiv y, x_1\equiv x$. With $A_{y}(z)\equiv \phi(z)$ we obtain
\bea
e^{2\rho} X^{abcd}F_{ab}F_{cd}&=& 4 X^{zy}_{\ \ zy} g^{zz}\phi_{\omega}'(z) \phi_{-\omega}'(z) \nonumber \\
e^{2\rho} X^{abcd}n_a n_c e^{2\rho} F_{be}F_{d}^{\ e}&=& (X^{xz}_{\ \ xz}+X^{xy}_{\ \ xy}) g^{zz}\phi_{\omega}'(z) \phi_{-\omega}'(z).
\eea
Therefore the two derivative part of the effective action leads to
\be
\sqrt{-g}\left(-2 X^{zy}_{\ \ zy}+X^{xy}_{\ \ xy}\right)
\ee
where we have used $X^{xz}_{\ \ xz}=X^{yz}_{\ \ yz}$ by rotational invariance.
Next up are the four derivative terms. These are
\bea
\mathcal Y&\equiv &Y^{xabc,}_{\ xdef}e^{-2\rho} \nabla_a\left(e^{3\rho}F_{bc}\right)\nabla^d\left(e^{3\rho}F^{ef}\right) 
=4\left\{ Y^{xzzy}_{\ xzzy}e^{-6\rho} [\nabla_z(e^{3\rho}F_{zy})]^2\right.\nonumber \\
 &+&\left. 2Y^{xmny}_{\ xzzy}e^{-3\rho} \nabla_m F_{ny} \nabla_z(e^{3\rho}F_{zy})+ Y^{xmny}_{ \ xpqy}\nabla_{m}F_{ny}\nabla^{p}F^{qy} \right\}.
\eea
Here $m,n,p,q$ indices refer to all coordinates with the exception of $z$. For brevity we will now define $Y^{ab}\equiv Y^{xaay}_{xbby}g^{bb}g_{aa}$.
Taking the near horizon limit and keeping only the poles we obtain:
\bea
\frac{\mathcal Y}4&=& 6 \rho' (Y^{zz}-Y^{tz})(\nabla_{z}F^{zy})F^{z}_{\ y}+(Y^{tt}-Y^{zz})(\nabla_{t}F^{t}_{\ y})^2 \nonumber \\
&+& 2 (Y^{tz}-Y^{zz})(\nabla_{t}F^{t}_{\ y})(\nabla_{z}F^{z}_{\ y})-2(Y^{zx_i}-Y^{tx_i})(\nabla_{t}F^{t}_{\ y})(\nabla_{x_i}F^{x_i}_{\ y}).
\eea
The differences of $Y$ tensors correspond to the $\alpha$ parameters defined in \reef{alphas}.
In the above expression the second and third terms are actually double poles. This imposes the regularity condition $Y^{tt}=Y^{zz}$. The other term is harmless as the double pole arises from the presence of a total derivative. Here we simply take the simple pole part, as this is all our method requires. It is easy to show that
\bea
\nabla_{z}F^{z}_{\ y}&=& -\frac 12 e^{-2g}\left(1+2z g'+2z \rho'\right)\phi'(z) \nonumber \\
\nabla_{t}F^{t}_{\ y}&=& \frac 12 e^{-2g}\left(1+2z f'\right)\phi'(z) \nonumber 
\eea
The single pole part is extracted as follows:
\bea
\sqrt{-g}\alpha^{zt}(\nabla_{t}F^{t}_{\ y})(\nabla_{z}F^{z}_{\ y})
\simeq \mathcal O(1/z^2)+\frac 14 \sqrt{-g}e^{-2g}
\left[\partial_z \alpha^{zt}-\alpha^{zt}e^{2g}R\right](g^{zz}\phi'(z) \phi'(z))
\eea
where we have used $$\lim_{z\to 0}R=-e^{-2g}(3f'-g'+2(d-1)\rho')$$.
Putting all the ingredients together the effective two derivative lagrangian reads
\be
\mathcal L^{(2)}=\frac{V}{128 \pi^2 T G_N}\frac{\omega^2}{ z}
\left(2 X^{zy}_{~~zy}-X^{xy}_{~~xy}-4 e^{-2g_0}\left[\partial_z \alpha^{zt}-\alpha^{zt}\left(e^{2g_0} R+(d+3)\rho'_0-2 \sum_i\alpha^i \rho_0'\right)\right]\right)
\ee
Using \reef{PoleRadialFormula} then leads to the result \reef{RadialFormula}, except for the $\alpha^i$ terms, defined as
\be
\alpha^i=Y^{zx_i}-Y^{tx_i}.\ee
These are actually zero by regularity. This follows from deducing the time formula \reef{TimeFormula}, which can be done in an analogous way to what we have just done. However, in this deduction the $\alpha^i$ coefficients do not enter. Equality of the two formulae then implies the constraints:
\bea
Y^{xz~~,yz}_{~~~zy~~~zx}-Y^{xz~~,yt}_{~~~zy~~~tx}&=& Y^{xt~~,yz}_{~~~zy~~~tx}+Y^{xt~~,yt}_{~~~zy~~~zx} \nonumber, \\
Y^{xz~~,yz}_{~~~zy~~~zx}-Y^{xt~~,yt}_{~~~ty~~~tx}&=&0 \nonumber \\
Y^{xx_i~~,yz}_{~~~x_i y~~~z x}-Y^{xx_i~~,y t}_{~~~x_iy~~~tx}&=& 0 \nonumber \\
X^{ty}_{~~ty}-X^{zy}_{~~zy}&=&0.\label{regconstraints}
\eea
The equalities above should be true for any background and lagrangian, sufficiently close to the horizon.

\section{Functions in 4.12}
The functions in equation \reef{eqnfunctions} are given by
\bea
\frac{A(z)}{\sqrt{-g}g^{zz}}&=& 
1-64 \gamma(18-58 z+77z^2-48 z^3+12 z^4) \nonumber \\
\frac{B(z)}{\sqrt{-g}g^{zz}}&=&
64\gamma (1-z)(24-106 z+177z^2-124 z^3+31z^4) \nonumber \\
\frac{C(z)}{\sqrt{-g}g^{zz}}&=&
-192\gamma (1-z)^2(4-22z+47z^2-36 z^3+9z^4)\nonumber \\
\frac{D(z)}{\sqrt{-g}(g^{zz})^2}&=&
96\gamma (1-z)^2 \nonumber \\
\frac{E(z)}{\sqrt{-g}(g^{zz})^2}&=&
-32\gamma (1-z)(3-10 z+5 z^2)\nonumber \\
\frac{F(z)}{\sqrt{-g}(g^{zz})^3}&=&-4 \gamma \nonumber
\eea

\bibliography{WaldPaperV2}{}
\bibliographystyle{JHEP}

\end{document}